\documentclass[a4paper,12pt]{amsart}
\usepackage{hyperref,latexsym}
\usepackage{enumerate}

\theoremstyle{plain}
\newtheorem{theorem}{Theorem}%[section]
\newtheorem{corollary}{Corollary}
\newtheorem{proposition}[theorem]{Proposition}
\theoremstyle{definition}

\theoremstyle{remark}

\begin{document}

\title[New Semi--Hamiltonian hierarchy]
      {New Semi--Hamiltonian hierarchy related to integrable magnetic flows on surfaces}

\date{13 November 2011}
\author{Misha Bialy and Andrey Mironov}
\address{M. Bialy, School of Mathematical Sciences, Raymond and Beverly Sackler Faculty of Exact Sciences, Tel Aviv University,
Israel} \email{bialy@post.tau.ac.il}
\address{A.E. Mironov, Sobolev Institute of Mathematics and
Laboratory of Geometric Methods in Mathematical Physics, Moscow State University }
\email{mironov@math.nsc.ru}
\thanks{M.B. was supported in part by Israel Science foundation grant 128/10 and A.M. was
supported by RFBR grant 11-01-12106-ofi-m-2011}


\begin{abstract}
We consider magnetic geodesic flows on the 2-torus. We prove that
the question of existence of polynomial in momenta first integrals
on one energy level leads to a Semi-Hamiltonian system of
quasi-linear equations,
 i.e. in the
hyperbolic regions the system has Riemann invariants and can be written in conservation laws form.

\end{abstract}

\maketitle

\section{Introduction}
In this paper we introduce a new hierarchy of Semi-Hamiltonian
system which is naturally related to integrable magnetic flows on
surfaces.

More precisely we consider magnetic geodesic flows on two-torus.
Consider one energy level and assume it admits a polynomial in
momenta integral of motion. Then we prove that the system of
quasi-linear equations on the coefficients is in fact
Semi-Hamiltonian system. These systems were introduced by S.Tsarev
and later on studied extensively. It  is proved in \cite{Ts} that
these systems are integrable by the generalized hodograph method. A
remarkable theorem by Sevennec \cite{sev} states that
Semi-Hamiltonian property is equivalent to existence of two special
coordinate systems in the space of field variables: Riemann
invariants and Conservation laws. It is remarkable fact that both
forms naturally appear for many systems. For example for Benney
chains \cite{B0} and for geodesic flows \cite{BM2} the Riemann
invariants correspond to critical values of the integral, and
Conservation laws are related to the invariant torii of the flow. In
this paper the Semi-Hamiltonian property appears naturally in the
same manner as we shall prove below.

The problem of existence of integrable systems in the presence of
gyroscopic forces (which is equivalent to magnetic field) was
studied by V.V.Kozlov and his students \cite{K},\cite{T},\cite{B}.
Topological obstructions to the integrability of mechanical systems
on surfaces with magnetic fields were obtained in \cite{B}.

First of all let us recall some facts about geodesic flows on
2-torus (without magnetic field). If the geodesic flow is integrable
then on the torus there are global semi-geodesic coordinates
$ds^2=g^2(t,x)dt^2+dx^2$. This coordinates can be constructed using
invariant Liouville torus having the diffeomorphic projection on the
base $\mathbb{T}^2$. The existence of such a torus is proven in
\cite{B1}. In the semi-geodesic coordinates the polynomial in
momenta integral has the form
$$
 F_n=\sum_{k=0}^na_{k}(t,x)\frac{p_1^{n-k}}{g^{n-k}}p_2^{k},
$$
where $a_{n-1}\equiv g \quad and \quad a_n\equiv 1.$ The coefficients satisfies the system
$$
 U_t+A(U)U_x=0,\eqno(1)
$$
where $U=(a_0,\dots,a_{n-2},a_{n-1})^{T},$ and matrix $A$ has the form
$$
 A=  \left(
  \begin{array}{cccccc}
   0 & 0 & \dots &
  0 & 0 & a_1  \\
  a_{n-1} &
 0 & \dots & 0 & 0 & 2a_2-na_0\\0 &
 a_{n-1} & \dots & 0 & 0 & 3a_3-(n-1)a_1\\
 \dots & \dots & \dots & \dots & \dots & \dots \\
 0 &
 0 & \dots & a_{n-1} & 0 & (n-1)a_{n-1}-3a_{n-3}\\
 0 &
 0 & \dots & 0 & a_{n-1} & na_n-2a_{n-2}\\  \end{array}\right).
$$
In \cite{BM1} it is proven that in the case of integrals of degree 3 or 4 in the elliptic regions (where matrix $A$  has all eigenvalues
different and two of them are complex-conjugate) the integral can be reduced  to integrals of the first or second degree or the metric is flat.

We proved in \cite{BM2} that the system (1) is Semi-Hamiltonian. In
this paper we generalize this result to the case of nontrivial
magnetic field.

\section{The main theorem}
The geodesic flow on the torus ${\mathbb T}^2$ with the Riemannian metric $ds^2=g_{ij}dx^idx^j$ given by the Hamiltonian
equations on $T^*{\mathbb T}^2$
$$
 \dot{x}^j=\frac{\partial H}{\partial p_j},\  \dot{p}_j=-\frac{\partial H}{\partial
 x^j},\ j=1,2,
$$
where $H=\frac{1}{2}g^{ij}p_ip_j$.
The function $F:T^*{\mathbb T}^2\rightarrow {\mathbb R}$ is called the first integral of the geodesic flow if
$$
 \dot{F}=\{F,H\}_g=0,
$$
where $\{F,H\}_g$ is Poisson bracket
$$
 \{F,H\}_g=\sum_{j=1}^2\left(\frac{\partial
 F}{\partial x^j}\frac{\partial
 H}{\partial p_j}-\frac{\partial
 H}{\partial x^j}\frac{\partial
 F}{\partial p_j}\right).
$$
In the case of magnetic geodesic flow the Poisson bracket gets the
form
$$
 \{F,H\}_{mg}=\sum_{j=1}^2\left(\frac{\partial
 F}{\partial x^j}\frac{\partial
 H}{\partial p_j}-\frac{\partial
 H}{\partial x^j}\frac{\partial
 F}{\partial p_j}\right)+\Omega\left(\frac{\partial F}{\partial p_1}\frac{\partial H}{\partial p_2}-
 \frac{\partial F}{\partial p_2}\frac{\partial H}{\partial p_1}\right),
$$
where $\Omega$ is the magnetic field. The magnetic geodesic flow given by the Hamiltonian equations with respect to
magnetic geodesic bracket $\{.,.\}_{mg}$
$$
 \dot{x}^i=\{x^i,H\}_{mg},\ \dot{p}_i=\{p_i,H\}_{mg}. \eqno{(2)}
$$
Let us choose conformal coordinates $(x,y)$ on the torus. The metric becomes the form $ds^2=\Lambda(x,y)(dx^2+dy^2).$

We shall consider the problem of existence of polynomial in momenta
integral  of motion $F$ of degree $N$ on one energy level
$H=\frac{p_1^2+p_2^2}{2\Lambda}=\frac{1}{2}$. We can parameterize
the fibres of the energy level over $\mathbb{T}^2$ as follows. Write
$$
 p_1=\sqrt{\Lambda}\cos\varphi,\ p_2=\sqrt{\Lambda}\sin\varphi.
$$
The equations (2) becomes of the form
$$
 \dot{x}=\frac{1}{\sqrt{\Lambda}}\cos\varphi,\ \dot{y}=\frac{1}{\sqrt{\Lambda}}\sin\varphi,
$$
$$
 \dot{\varphi}=\frac{\Lambda_y}{2\Lambda\sqrt{\Lambda}}\cos\varphi-\frac{\Lambda_x}{2\Lambda\sqrt{\Lambda}}\sin\varphi-\Omega.
$$
The integral $F$ on the energy level has the form
$$
 F=\sum_{k=-N}^{k=N}a_ke^{ik\varphi},\eqno{(3)}
$$
where $a_k=u_k(x,y)+iv_k(x,y),\ a_{-k}=\bar{a}_k$.

The condition $\dot{F}=0$ is equivalent to

$$
 (F)_x\cos\varphi+(F)_y\sin\varphi+
 F_{\varphi}\left(\frac{\Lambda_y}{2\Lambda}\cos\varphi-\frac{\Lambda_x}{2\Lambda}\sin\varphi-\Omega\sqrt{\Lambda}\right)=0.
$$

We substitute (3) in the last equation and collect terms with
respect to $e^{ik\varphi}$. We get

$$
 \frac{(a_{k-1})_x+(a_{k+1})_x}{2}+\frac{(a_{k-1})_y-(a_{k+1})_y}{2i}+\frac{\Lambda_y}{2\Lambda}\frac{i(k-1)a_{k-1}+i(k+1)a_{k+1}}{2}-
$$

$$
 -\frac{\Lambda_x}{2\Lambda}\frac{i(k-1)a_{k-1}-i(k+1)a_{k+1}}{2i}-ik\Omega\sqrt{\Lambda}a_k=0,\eqno{(4)}
$$

where $k=0,\dots,N+1$ (we assume $a_s=0$ at $s>N$).

Let $k$ be $N+1$ in (4):
$$
 (a_N)_x-N\frac{\Lambda_x}{2\Lambda}a_N+\frac{1}{i}\left((a_N)_y-N\frac{\Lambda_y}{2\Lambda}a_N\right)=0.
$$

Multiply the last identity by $\Lambda^{-N/2}$

$$
 (a_N\Lambda^{-N/2})_x-i(a_N\Lambda^{-N/2})_y=0.
$$

Thus $a_N\Lambda^{-N/2}$ is a holomorphic function, consequently
$a_N=\Lambda^{N/2}(\alpha+i\beta)$ for some constants
$\alpha,\beta$. Taking appropriate rotation in the plane $(x,y)$ and
dividing $F$ by appropriate constant we can assume that
$\alpha=1,\beta=0$. Thus we have $a_N=\Lambda^{N/2}$.

Notice that for $k=0$ the equation (4) does not contain magnetic
field and has the form

$$
 \frac{(a_{-1})_x+(a_{1})_x}{2}+\frac{(a_{-1})_y-(a_{1})_y}{2i}+\frac{\Lambda_y}{2\Lambda}\frac{ia_{1}-ia_{-1}}{2}+
 \frac{\Lambda_x}{2\Lambda}\frac{a_{-1}+a_1}{2}=0.\eqno{(5)}
$$

Let us introduce the notation
$$
 Q_k=
 \frac{(a_{k-1})_x+(a_{k+1})_x}{2}+\frac{(a_{k-1})_y-(a_{k+1})_y}{2i}+
$$

$$
 +\frac{\Lambda_y}{2\Lambda}\frac{i(k-1)a_{k-1}+i(k+1)a_{k+1}}{2}-\frac{\Lambda_x}{2\Lambda}\frac{(k-1)a_{k-1}-(k+1)a_{k+1}}{2}.
$$
From equation (4) for $k=N$ we can find the magnetic field:
$$
 \Omega=\frac{Q_N}{iN\sqrt{\Lambda}a_N}.
$$
Let us substitute this expression of $\Omega$ into (4) for every
$k=1,\dots,N-1$. We get
$$
 NQ_k\Lambda^{N/2}=kQ_Na_k.\eqno{(6)}
$$
The equations (5) and (6) form a system of quasi-linear equations on
 $2N$ unknown functions $\Lambda,u_0,\dots,$ $u_{N-1},$
$v_1,\dots,v_{N-1}.$ This is a quasi-linear system of the form
$$
 A(U)U_x+B(U)U_y=0,
$$
where  $U=(\Lambda,u_0,\dots,$ $u_{N-1},v_1,\dots,v_{N-1})^{\top}$.
We shall write in the last section this system explicitly for $N=2$.
Our main result is
\begin{theorem} {\it For any $N$, the quasi-linear system (5),(6)
on coefficients of the integral $F$  is Semi-Hamiltonian system.}
\end{theorem}

\section{Riemann invariants}
Consider $F$ as a function on the unite circle $S^1\subset{\mathbb
C}.$ It is a remarkable fact that the critical values of
$$
 F=\Lambda^{N/2}z^N+a_{N-1}z^{N-1}+\dots+\Lambda^{N/2}z^{-N}
$$
 are Riemann invariants of the system (5), (6). Indeed, let
$x_1,\dots,x_{2N}$ be the set of distinct critical points
(hyperbolic region) $x_i\in S^1\subset{\mathbb C}$. Introduce
$r_k=F(x_k), k=1,\dots,2N$. From the identity
$$
 \dot{F}=F_x\dot{x}+F_y\dot{y}+F_{\varphi}\dot{\varphi}=0
$$
it follows that $r_k$ are Riemann invariants, because having
$F_\varphi=0$ we are left with $F_x cos\varphi+ F_y sin\varphi=0$
and thus the equation on $r_k$ follows:

$$
 (r_k)_x+\lambda_k(r_k)_y=0,\ \lambda_k=\tan\varphi_k,\ x_k=e^{i\varphi_k}.
$$
Let us check that $r_1,...,r_{2N}$ form a regular coordinate system,
that is Riemann invariants are functionally independent. Write
$$
 zF'=N\Lambda^{N/2}z^N+(n-1)a_{n-1}z^{N-1}+\dots+a_1z-a_{-1}z^{-1}-\dots-N\Lambda^{N/2}z^{-N}.
$$
Notice that the critical points $x_1,\dots, x_{2N}$ are roots of
$zF'$ and so by Vieta formula
$$
 x_1\dots x_{2N}=-1.\eqno{(7)}
$$
For convenience we denote field variables as
$$
 (\mu_1,\dots,\mu_{2N})=(\Lambda^{N/2},a_{N-1},\dots,a_0,\dots,a_{1-N}).
$$
Then
$$
 \frac{\partial r_k}{\partial \mu_s}=\frac{\partial F}{\partial\mu_s}(x_k)+F'(x_k)\frac{\partial x_k}{\partial \mu_s}=
 \frac{\partial F}{\partial\mu_s}(x_k).
$$
Using (7) we have
$$
 \det\left(\frac{\partial r_k}{\partial \mu_s}\right)=(-1)^{N}
 \det
  \left(
  \begin{array}{cccc}
   x_1^{2N}+1 & x_1^{2N-1} & \dots &
  x_1 \\
  \dots &  \dots & \dots & \dots\\
  x_{2N}^{2N}+1 & x_{2N}^{2N-1} & \dots &
  x_{2N}\\  \end{array}\right).
$$
Splitting the first column and again using (7) we get
$$
 \det\left(\frac{\partial r_k}{\partial \mu_s}\right)=(-1)^N(x_1\dots x_{2N}-1)W=(-1)^{N+1}2W.
$$
Where $W=\prod_{i>j}(x_i-x_j)$ is the Vandermonde determinant. So we
have that $(\mu)\leftrightarrow(r)$ is a regular change of variable
near every point $A$ in the strictly hyperbolic region.

\noindent{\bf Remark.} The field variables $\Lambda,a_0$ are real,
$a_s$ and $a_{-s}, s>0$ are complex conjugate. Therefore computation
of $\left(\frac{\partial r_k}{\partial\mu_s}\right)$ is equivalent
to the computation of
$
 \frac{\partial(r_1,\dots,r_{2N})}{\partial(\Lambda^{N/2},a_0,u_1,\dots,u_{N-1},v_1,\dots,v_{N-1})}.
$

\section{Conservation laws}
The aim of this section is to prove that the system (5),(6) can be
written in the form of conservation laws. This system has many
explicit conservation laws. For example, the real part of (4) for
$k=N$ multiplied by $\Lambda^\frac{1-N}{2}$ has the form of
conservation low
$$
 \left(u_{N-1}\Lambda^{\frac{1-N}{2}}\right)_x+\left(v_{N-1}\Lambda^{\frac{1-N}{2}}\right)_y=0.
$$
Another series of conservation laws can be obtained in the following
way. The identity (4) at $k=0$ gives us a conservation law
$$
 (\sqrt{\Lambda}u_1)_x-(\sqrt{\Lambda}v_1)_y=0.
$$
Similarly we can get this conservation law for the powers of the
integral. Namely $F^m$ generates the conservations law

$$
 (\sqrt{\Lambda}u_1^{(m)})_x-(\sqrt{\Lambda}v_1^{(m)})_y=0.
$$
where $a_{1}^{(m)}=u_{1}^{(m)}+i v_{1}^{(m)}$ are corresponding
Fourier coefficients of $F^m$. So, we have infinitely many explicit
conservation lows. Remarkably they are in fact polynomial in the
field variables. However we do not know if it is possible to get by
this method functionally independent conservation lows. For this
reason we proceed in a different way. We show that is possible to
generate functionally independent conservation laws by invariant
tori of the magnetic flow, in a similar way we did it in \cite {BM2}
for geodesic flows.

Imaginary part of (4) for $k=N$ multiplied by
$\Lambda^\frac{1-N}{2}$ gives us
$$
 \Omega\Lambda=\frac{1}{2N}\left(v_{N-1}\Lambda^{(1-N)/2}\right)_x-\frac{1}{2N}\left(u_{N-1}\Lambda^{(1-N)/2}\right)_y. \eqno{(8)}
$$
Let $\varphi=f(x,y)$ be a surface invariant under the flow. The
invariance condition reads as follows:
$$
 \frac{f_x\cos f}{\sqrt{\Lambda}}+\frac{f_y\sin f}{\sqrt{\Lambda}}+\frac{\Lambda_x\sin f}{2\Lambda\sqrt{\Lambda}}
 -\frac{\Lambda_y\cos f}{2\Lambda\sqrt{\Lambda}}+\Omega=0
$$
or equivalently multiplying by $\Lambda$
$$
 (\sqrt{\Lambda}\sin f)_x-(\sqrt{\Lambda}\cos f)_y+\Omega\Lambda=0.\eqno{(9)}
$$
Substituting (8) into (9) we get a conservation law corresponding to
invariant surface $\varphi=f(x,y)$:
$$
 \left(\sqrt{\Lambda}\sin f+\frac{1}{2N}v_{N-1}\Lambda^{(1-N)/2}\right)_x-\left(\sqrt{\Lambda}\cos f+\frac{1}{2N}u_{N-1}\Lambda^{(1-N)/2}\right)_y=0.
$$
Let us show now that by this method we can get, in the strictly
hyperbolic region, $2N$ conservation laws with functionally
independent $G_k$,
$$
 G_k={\rm Im}[\sqrt{\Lambda}z_k+\frac{1}{2N}a_{N-1}\Lambda^{(1-N)/2}],\ z_k=e^{i\varphi_k}.
$$
Let $\Gamma$ be the domain of strict hyperbolicity of $F$ ($F'$ has
$2N$ distinct roots on the unite circle).

\begin{proposition} {\it Denote by $\hat{\Gamma}$ the open dense subset of $\Gamma$ defined by the condition that $\pm i$ is not among
the roots of $F$. Let $A^*=(\mu_1^*,\dots,\mu_{2N}^*)$ be any point
of $\hat\Gamma$. Then in a neighborhood of $A^*$ there exist $2N$
functionally independent conservation laws. }
\end{proposition}

\begin{corollary} {\it Quasi-linear system (5),(6) is Semi-Hamiltonian at any point of $\Gamma$.}
\end{corollary}

\noindent{\bf Proof of Corollary.}
 By Sevennec theorem and the
Proposition 1 the system is Semi-Hamiltonian at any point of
$\hat{\Gamma}$. But it is dense, so the condition of
Semi-Hamiltonicity extends to the whole $\Gamma$. Applying Sevennec
theorem again we have that also for
$A^*\in\Gamma\backslash\hat{\Gamma}$ there are $2N$ independent
conservations laws, but we were not able to construct them
explicitly. This proves the Corollary.  $\square$

\noindent{\bf Proof of Proposition.} Define a neighborhood of $A^*$
with the help of Riemann invariants $r_1,\dots,r_{2N}$. Recall $r_k$
are local coordinates.

Let $r_k^*,k=1,\dots,2N$ are Riemann invariants of $A^*$. That is
$x^*_k$ are critical points of
$F^{*}=(\Lambda^*)^{N/2}z^N+a^*_{N-1}z^{N-1}+\dots+(\Lambda^*)^{N/2}z^{-N}$.
Notice that all critical points $x^*_k$ are non-degenerate since we
are in the strictly Hyperbolic region. We shall assume that odd and
even values of the index $k$ correspond to  maxima and minima
respectively.
 Let $\varepsilon>0$ be a small
number. Define a neighborhood of $F^*$
$$
 R_{\varepsilon}=\{F: |r_k-r_k^*|<\frac{\varepsilon}{10},k=1,\dots,2N \}.
$$
Here $\varepsilon$ should be chosen smaller then $\frac{1}{4}{\rm
min}|r_k^*-r^*_{k+1}|$.

So for any polynomial from $R_{\varepsilon}$ it has critical points
$x_k$ close to $x_k^*$ on the unite circle and critical values close
to $r_k^*$. Define now $z_k(\varepsilon,\mu)$ as solutions of the
equations on the unite circle:
$$
 F(z_k(\varepsilon,\mu),\mu)=r_k+(-1)^k\varepsilon,
$$
$z_k$ lies in a neighborhood of $x_k$. Since we assume that
$x_1,x_3,\dots$ are points of maxima and $x_2,x_4,\dots$ are points
of minima then it follows that $r_k+(-1)^k\varepsilon$ are regular
values for any polynomial $F$ with coefficients taken from
$R_{\varepsilon}$ ($F$ should be restricted to a neighborhoods of
$x_k^*$ ). Important for us is that $z_k(\varepsilon,\mu)$ depend
smoothly on $\varepsilon$ and $\mu$. When we shrink $\varepsilon$ to
zero then $z_k(\mu,\varepsilon)\rightarrow x_k$ by the construction.
In the following we shall decrease $\varepsilon$, but keeping away
from zero, in order to get the needed neighborhood
$R_{\varepsilon}$.

Define the following functions on $R_{\varepsilon}$:
$$
 G_k={\rm Im}[\sqrt{\Lambda}z_k+\frac{1}{2N}a_{N-1}\Lambda^{(1-N)/2}]=
$$
$$
 \frac{1}{2i}
 {\rm Im}\left[\sqrt{\Lambda}\left(z_k-\frac{1}{z_k}\right)+\frac{1}{2N}\Lambda^{(1-N)/2}(a_{N-1}-a_{1-N})\right]
$$
(where $z_k$ depend on $\mu$s implicitly). We claim that one can
choose $\varepsilon>0$ sufficiently small in order to have
$$
 {\rm det}\frac{\partial(G_1,\dots,G_{2N})}{\partial(\mu_1,\dots,\mu_{2N})}(A^*)\ne 0.
$$
This would imply the claim.
We have
$$
 \frac{\partial G_k}{\partial\mu_l}|_{\mu=\mu^*}=-\frac{\sqrt{\Lambda^*}}{2i}\left(1+\frac{1}{(z_k^*)^2}\right)
 \frac{\frac{\partial F}{\partial \mu_l}|_{\mu=\mu^*}}{F'(\mu^*,z_k)}+R^*_{kl}.
$$
Where $R^*_{kl}$ contains all terms of explicit derivation of
$\frac{\partial G_k}{\partial \mu_l}$.

We have
$$
 {\rm det}\left(\frac{\partial G}{\partial\mu}\right)|_{\mu=\mu^*}=\left(\frac{\sqrt{\Lambda^*}}{2i}\right)^{2N}\prod_{k=1}^{2N}\left(1+\frac{1}{(z_k^*)^2}\right)
 \prod_{k=1}^{2N}\frac{1}{F'(\mu^*,z_k)}\times
$$
$$
 \times{\rm det}\left[\left(\frac{\partial F(z_k^*,\mu)}{\partial\mu_l}\right)|_{\mu=\mu^*}+R_{kl}^*\left(\frac{1}{1+\frac{1}{(z_k^*)^2}}\right)F'(\mu^*,z_k^*)\right].
$$
Mention that  when $\varepsilon\rightarrow 0$, $\frac{1}{1+\frac{1}{(z_k^*)^2}}\rightarrow\frac{1}{1+\frac{1}{(x_k)^2}}$ which is finite
since $x_k\ne i$ by assumptions. Also $z_k^*(\varepsilon)\rightarrow x_k^*$ and so $F'(\mu^*,z_k^*)\rightarrow 0$.
Therefore the determinant in brackets when $\varepsilon\rightarrow 0$ tends to the determinant of the matrix
$$
 \left(\frac{\partial F(x_k^*,\mu)}{\partial\mu_l}\right).
$$
This is exactly the determinant of the matrix considered in the
section 3. So it is equal $-2W(x_1^*,\dots,x_{2N}^*)$ and does not
vanish. But then it follows that for small $\varepsilon>0$ ${\rm
det}\left(\frac{\partial G}{\partial \mu}(A)\right)\ne 0$ in a
neighborhood of $A^*$. Proposition 2 and Theorem 1 are proved.

\section{Discussion}
In this section we discus some open problems.

In the case of $n=2$ the equations (5),(6) on functions $\Lambda, u_0,u_1,v_1$ have the form
$$
 (\sqrt{\Lambda}u_1)_x-(\sqrt{\Lambda}v_1)_y=0,
$$
$$
 \left(\frac{u_1}{\sqrt{\Lambda}}\right)_x+\left(\frac{v_1}{\sqrt{\Lambda}}\right)_y=0,
$$
$$
 (u_0)_x+2\Lambda_x-\frac{v_1}{2\sqrt{\Lambda}}\left(\left(\frac{u_1}{\sqrt{\Lambda}}\right)_y-\left(\frac{v_1}{\sqrt{\Lambda}}\right)_x\right)=0,
$$
$$
 -(u_0)_y+2\Lambda_y-\frac{u_1}{2\sqrt{\Lambda}}\left(\left(\frac{u_1}{\sqrt{\Lambda}}\right)_y-\left(\frac{v_1}{\sqrt{\Lambda}}\right)_x\right)=0.
$$
Introduce the new functions $f,g:$
$$
 f=\frac{u_1}{\sqrt{\Lambda}},\ g=\frac{v_1}{\sqrt{\Lambda}}.
$$
We have
$$
 f_x+g_y=0,
$$
$$
 (f\Lambda)_x-(g\Lambda)_y=0,
$$
$$
 (u_0)_x+2\Lambda_x-\frac{1}{2}g(f_y-g_x)=0,
$$
$$
 -(u_0)_y+2\Lambda_y+\frac{1}{2}f(f_y-g_x)=0.
$$
 This system can be written in the form
$$
 A(U)U_x+B(U)U_y=0,
$$
where $U=(\Lambda,u_0,f,g)$,
$$
 A=  \left(
  \begin{array}{cccc}
   0 & 0 & 1 & 0 \\
  f & 0 & \Lambda & 0 \\
   2 & 1 & 0 & \frac{1}{2} g \\
 0 & 0 & 0 & -\frac{1}{2}f \\
\end{array}\right), \
B=  \left(
  \begin{array}{cccc}
   0 & 0 & 0 & 1 \\
  -g & 0 & 0 & -\Lambda \\
   0 & 0 & -\frac{1}{2}g & 0 \\
  2 & -1 & \frac{1}{2}f & 0 \\
\end{array}\right).
$$
The problem of existence of periodic solution is very interesting.
The systems of such form (non-evolution form) were considered in
\cite{B2} from the point of view of blow-up analysis along
characteristic curves. It would be very interesting to apply these
ideas to our system.

Notice
$$
 gf_y=(fg)_y-fg_y=(fg)_y+ff_x=0,
$$
$$
 fg_x=(fg)_x-gf_x=(fg)_x+gg_y.
$$
So we have
$$
 (u_0)_x+2\Lambda_x+\frac{1}{2}(gg_x-ff_x)-\frac{1}{2}(fg)_y=0,
$$
$$
 -(u_0)_y+2\Lambda_y+\frac{1}{2}(ff_y-gg_y)-\frac{1}{2}(fg)_x=0.
$$
Thus we have explicit conservation laws form for our system
$$
 f_x+g_y=0,
$$
$$
 (f\Lambda)_x-(g\Lambda)_y=0,
$$
$$
 (u_0+2\Lambda+\frac{1}{4}(g^2-f^2))_x-(\frac{1}{2}fg)_y=0,\eqno{(10)}
$$
$$
 (-u_0+2\Lambda-\frac{1}{4}(g^2-f^2))_y-(\frac{1}{2}fg)_x=0.\eqno{(11)}
$$
Lust two conservation lows are very interesting.
Let us recall that a hyperbolic diagonal system
$$
 (r_i)_x+\lambda_i(r_1,\dots,r_n)(r_i)_y=0, i=1,..,n
$$ is Semi-Hamiltonian if
$$
 \partial_j\Gamma^k_{ki}=\partial_i\Gamma^k_{kj},\ i\ne j\ne k,
$$
where
$$
 \Gamma^k_{ki}=
\frac{\partial_{i}\lambda_k}{\lambda_i-\lambda_k}.
$$
It can be proved that a diagonal system is Semi-Hamiltonian if and
only if it can be written in some coordinates as a system of
conservation laws (see \cite{sev}). We have the diagonal metric (see
\cite{Ts}) $g_{ii}=H_i^2$, and Lame coefficients can be found from
the over-determined system
$$
 \partial_k\ln H_i=\Gamma^i_{ik}.
$$
By Pavlov--Tsarev theorem \cite{PT} if the Semi-Hamiltonian system has two conservation lows of the form
$$
 F_x+G_y=0,\ F_y+H_x=0,
$$
then the corresponding metric $g_{ii}$ is Egorov metric i.e.
$$
 \beta_{ij}=\beta_{ji}, \ \beta_{ij}=\frac{\partial_iH_j}{H_i},\ i\ne j.
$$
In such a case the metric is potential: $g_{ii}=\partial_i a(r)$ for
a function $a$. So it follows that our system for $N=2$ is Egorov
Semi-Hamiltonian system, since we have two conservation lows
(10),(11). By similar calculations one can check that for $N=3$ our
system is also Egorov system. It would be interesting to prove this
fact for arbitrary $N$.

Another interesting problem is to find Poisson bracket of hydrodynamical type for the system in the form of Dubrovin--Novikov \cite{DN}
or Ferapontov--Mokhov \cite{FM}.

\end{document}